\documentclass[aps,preprint]{revtex4}%
\usepackage{amsfonts}
\usepackage{amsmath}
\usepackage{amssymb}
\usepackage{graphicx}%

\begin{document}
\title[Features of PEL in BLJ mixtures]{Potential Energy Landscape in Lennard-Jones binary mixture model}
\author{M. Sampoli}
\affiliation{Dipartimento di Energetica and INFM, Universit\`{a} di Firenze Via Santa Marta
3, I-5019, Firenze, Italy}
\author{P. Benassi, R. Eramo}
\affiliation{Dipartimento di Fisica and INFM, Universit\`{a} di L'Aquila, Via Vetoio,
Coppito, Aquila, Italy}
\author{L. Angelani, G. Ruocco}
\affiliation{Dipartimento di Fisica and INFM, Universit\`{a} di Roma \textquotedblleft La
Sapienza\textquotedblright, P.le Aldo Moro 2, I-00185, Roma, Italy}

\begin{abstract}
The potential energy landscape in the Kob-Andersen Lennard-Jones binary
mixture model has been studied carefully from liquid down to the supercooled
regime, from T $=2$ down to T $=0.46$. One thousand of independent
configurations along the time evolution have been examined at each
investigated temperature. From the starting configuration we searched the
nearest saddle (or quasi-saddle) and minimum of the potential energy. The
vibrational densities of states for the starting and the two derived
configurations have been evaluated. Besides the number of negative eigenvalues
of saddle, other quantities show some signature of the approaching of the
dynamical arrest temperature.

\end{abstract}
\date{29 December 2002}
\startpage{1}
\maketitle

\section{Introduction}

In recent years a considerable research effort has been devoted to understand
the complex phenomenology of supercooled glass forming liquids and in
particular the enormous increase of relaxation times and viscosity by many
order of magnitude upon decreasing the temperature. It was
Goldstein\cite{Goldstein69} who first related the behaviour of glass formers
to the underlying Potential Energy Landscape (PEL) and proposed to
characterize the dynamics of the system through the motion of a point in the
complex high-dimensional PEL ($3N$-dimensional if $N$ is the total number of
particles). Further, he suggested to focus onto the PEL local minima where the
system is supposed to be trapped at low enough temperatures. At low
temperatures, but above the glass transition, the phase space spanned by the
system at equilibrium can ideally be characterized by two different types of
processes: a "fast" relaxation into local minima (or basins) and the "slow"
relaxation due to the crossing between loosely connected basins (the so called
"hopping" from basin to basin even if, in the classical mechanics PEL
language, there exist only tortuous and crooked ways connecting two admissible
basins). Goldstein focused the attention on the minima of PEL, but before the
system is spending much of the time in a given basin, there are situations
where "fast" and "slow" relaxations are not recognizable, diffusion is still
high and a lot of minima are visited in a continuos way. In this case, it was
searched to relate the diffusion dynamics to unstable directions in
equilibrium (or instantaneous) PEL configurations, i.e. in the context of
Instantaneous Normal Mode approach\cite{Keyes97,Donati00}, however other
special points than PEL minima, are found to characterize better the dynamics
of the system: they are the minima of the modulus square of the gradient of
the potential energy, i.e. saddles (absolute minima) and what we call
"quasi-saddles" (local minima). Indeed the unstable directions of a saddle
give some piece of information about the number of different minima are
existing in that PEL region and therefore about the diffusion processes. In
the past, the relevance of saddles for the dynamics of glassy systems has been
already recognized in the context of mean-field
spin-glasses\cite{Kurchan96,Cavagna98}. A large number of recent works has
renew interest in PEL and its characterization through saddle and minima
points to better understand the continuos transition from liquid to
glass\cite{PRL85-5356, PRL85-5360, Shah01, Cavagna01, Grigera02}.

The purpose of the present work is to investigate in detail the statistical
properties of the PEL of a binary Lennard-Jones system from the liquid down to
the supercooled regime near the structural arrest. To do that, a set of
temperature above the glass transition have been considered and for each of
them a set of independent configurations has been examined. Nearest saddles
and minima have been searched and the normal mode analysis has been performed
on all configurations.

\section{Numerical computations}

The system under consideration is a binary mixture of classical particles. The
system is equal to that considered by Kob and Andersen\cite{Kob95}, apart from
a detail in the truncation of the potential. The mixture is made of $256$
particles, $205$ of $A$ and $51$ of $B$ (about $80\%$ of $A$ and $20\%$ of
$B$). Both particles have the same mass and interact each other via a
Lennard-Jones potential, i.e. $V_{\alpha\beta}\left(  r\right)  =4\epsilon
_{\alpha\beta}[(\sigma_{\alpha\beta}\diagup r)^{12}-$ $(\sigma_{\alpha\beta
}\diagup r)^{6}]$ with $\alpha$,$\beta\in\left\{  A,B\right\}  $. The
parameters of the potentials are: $\epsilon_{AA}=1.0$, $\sigma_{AA}=1.0$,
$\epsilon_{BB}=0.5$, $\sigma_{BB}=0.88$, $\epsilon_{AB}=1.5$, $\sigma
_{AB}=0.8$. In the following all the results will be given in reduced units,
i.e. length in units of $\sigma_{AA}$, energy in units of $\epsilon_{AA}$,
\ temperature in $k_{B}\diagup\epsilon_{AA}$, and time in units of
$(m_{A}\sigma_{AA}^{2}\diagup\epsilon_{AA})^{1/2}$.

The interaction potential at long distance is tapered between $0.95\ r_{2}%
=r_{1}\leq$ $r\leq r_{2}=2.56\ \sigma_{AA}$ with the following fifth-order
smoothing function $\mathcal{T}(r)=1+(r_{1}-r)^{3}(6r^{2}+(3r+r_{1}%
)(r_{1}-5r_{2})+10r_{2}^{2})\diagup(r_{2}-r_{1})^{5}$. In this way the
potential, the forces and their derivatives are continuos and it is possible
to keep the energy constant better than $1/10^{5}$ over $100$ millions of
molecular dynamics (MD) time steps near the critical temperature. The MD was
performed in NVE ensemble using the leapfrog algorithm with a time step of
$1.5\ 10^{-3}$ at temperature above $T=1.0$ and $2\ 10^{-3}$ at lower
temperatures. A neighbor-list was used to speed up the calculation. \ 

In their paper, Kob and Andersen\cite{Kob95} truncated and shifted the
potential at a cutoff distance of $2.5\ \sigma_{\alpha\beta}$. In the uniform
density approximation, our tapering correspond to an average cutoff distance
of about $2.5\ \sigma_{AA}$, \ and the resulting potential energy is shifted
(downward) only by about 0.56 energy units with respect to the values of the
quoted reference\cite{phishift}.

First the configurations were produced at high temperature and then
thermalized at the highest investigated temperature ($T=2.0$). The initial
configurations at lower temperatures were obtained in sequence by cooling and
thermalizing a set of different temperature ($1.8$, $1.6$, $1.4$, $1.2$,
$1.0$, $0.9$, $0.8$, $0.7$, $0.6$, $0.5$, $0.48$, $0.46$). Not all the
configurations were analyzed in full details.

For each temperature, $1000$ independent configurations were stored and then
analyzed. The minimum number of steps between two subsequent configurations to
be independent was determined in test runs. Two subsequent configurations are
classified as independent if different potential minima are always obtained
practically. We used a separation of $1000$ steps at high temperatures down to
$T=1.0$, then we increased this number following the viscosity trend, so we
used $10^{4}$ steps at $T=0.7$ and \ $10^{5}$ steps at $T=0.46$. The procedure
is similar to that used to study the inherent dynamics by Schroeder et
al.\cite{Schroeder00}

To characterize the states of the supercooled binary mixture, we searched the
nearest saddle point ("\emph{W}") and the nearest minimum ("\emph{V}") of the
PEL from the initial equilibrium ("\emph{I}") configurations. The saddle (or
quasi-saddle as explained after) configurations correspond to minimizing the
sum of the squared forces, i.e. $W=\left\vert \nabla V\right\vert ^{2}$ where
$V$ is the sum of all interaction potential of our system and $\nabla$\ is the
gradient on all the particle coordinates. If $\left\vert \nabla V\right\vert
^{2}$ is really vanishing, i.e. is an absolute minimum, the configuration
correspond to a real saddle, otherwise we call it quasi-saddle since usually
there is only one direction with zero or nearly zero eigenvalue and non zero
force\cite{cm0108310}. The minima of $V$ are always absolute minima of
$\left\vert \nabla V\right\vert ^{2}$.

To obtain good nearest configurations of $W$ and $V$ minimum from an initial
equilibrium configuration of our system of 256 particles is a very stiff
problem. The details of the two search procedure we used, is too long and will
be described elsewhere\cite{MS03}. Here we want to stress that various
standard algorithms have been employed during the search and various criteria
have been adopted for switching among the different algorithms. From the
numerical point of view, the obtained minima are quite satisfactory. Indeed at
the start, the value of $\left\vert \nabla V\right\vert ^{2}$ is of the order
of $10^{7}$ (in internal MD units) while at the minimum of $W$ is in the range
$10^{-1}\div10^{2}$ for quasi-saddles and in the range $10^{-8}\div10^{-2}$
for true-saddles. At the minimum of $V$ \ is always less than $10^{-3}$ (in
the range $10^{-8}\div10^{-2}$). In the worst case, the value of $\left\vert
\nabla V\right\vert ^{2}$ is reduced by a factor of $10^{5}$. This happens
when the flex point is on a high sliding surface.

%

\begin{figure}
[ptb]
\begin{center}
\includegraphics[
width=12cm
]%
{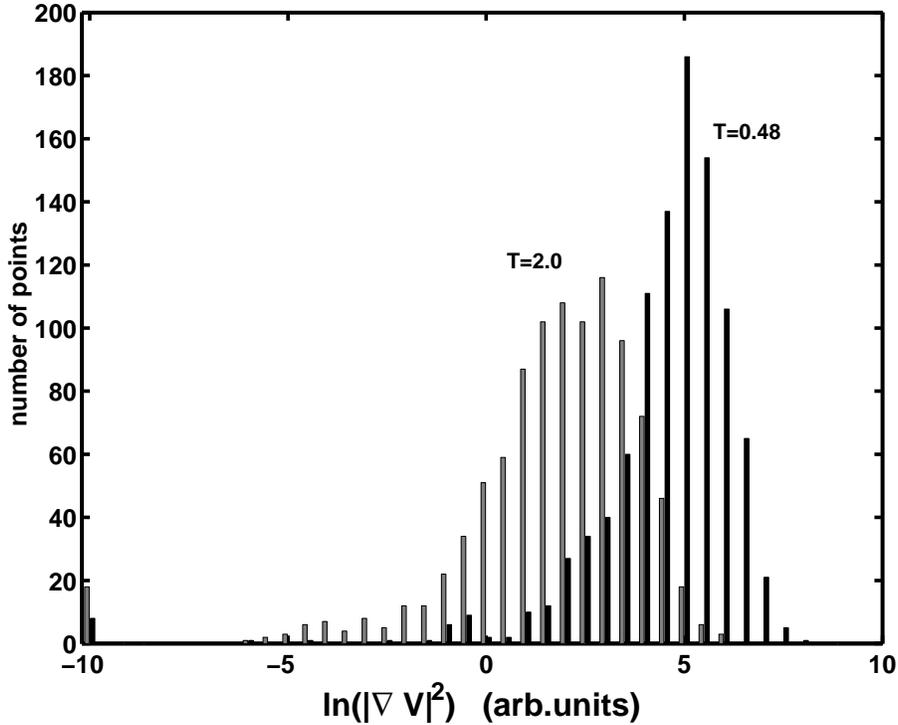}%
\caption{The distribution at high ($T=2.0$, black lines) and low ($T=0.48$,
grey lines) temperature of the $\left\vert \nabla V\right\vert ^{2}$\ values
for $W$ minimization. True-saddles are grouped around $-10$ value of ln(
$\left\vert \nabla V\right\vert ^{2}$).}%
\label{Adist_Sfqw}%
\end{center}
\end{figure}

In figure \ref{Adist_Sfqw} we plot the distribution of $\left\vert \nabla
V\right\vert ^{2}$ at the $W$ minima for two extreme values of temperature. It
should be noted that: i) the temperature strongly affects the distribution;
ii) the number of true-saddles (for which $\left\vert \nabla V\right\vert
^{2}$ is vanishing) is strongly decreasing with temperature; iii) the force
residuals, i.e. $\left\vert \nabla V\right\vert ^{2}$\ values, are larger at
low temperatures in contrast with what one can expect intuitively. Moreover
the temperature variation of the distribution is nearly confined to\ the low
temperature side (below $T=1.0$). The second point is the reason why it
becomes more and more difficult to analyze the low temperature data by using
only these configurations which have the $W$ minima as true-saddles. Therefore
we choose to analyze all the configurations and to inspect if true-saddles and
quasi-saddles share the same pieces of information.

\section{Analysis of the results}

\subsection{Behaviour \ of potential energy at minima and saddles}

%

\begin{figure}
[ptb]
\begin{center}
\includegraphics[
width=4.5463in
]%
{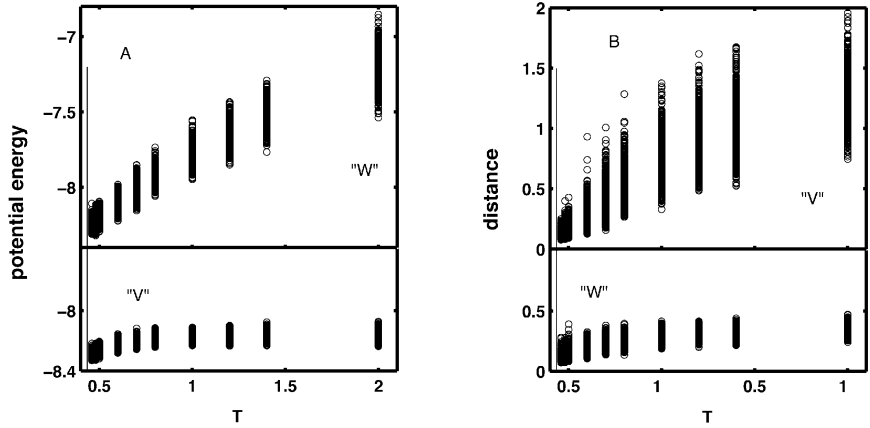}%
\caption{Panel A on the left. Potential energy (internal units) of
quasi-saddles ("\emph{W}",upper part) and potential minima ("\emph{V}", lower
part) versus temperture. Panel B on the right. Cartesian distance in box units
of potential minima ("\emph{V}", upper part) and quasi-saddles ("\emph{W}%
",lower part) from the starting equilibrium configuration versus temperature.
The thin vertical line on both the panels is drawn at T$_{c}$=0.435.}%
\label{Aphi_distVWa}%
\end{center}
\end{figure}

In figure \ref{Aphi_distVWa} we have reported the behaviour of the potential
energy versus temperature of the quasi-saddles ("\emph{W}") and minima
("\emph{V}"). As you can see, the potential energy of "\emph{W}" is decreasing
rapidly with temperature while that of "\emph{V}" is practically constant, as
already reported in ref. \cite{PRL85-5356}. Close and closer to the
temperature of structural arrest, even the "\emph{V}" potential energy becomes
a decreasing function of the temperature, as it has been already shown by
Sastry et al.\cite{Sastry99}

In the right panel of \ the same figure we have shown the Cartesian distances
(in box units) of "\emph{V}" and "\emph{W}" from the starting equilibrium
configurations. In that case the temperature behaviour of the two kind of
minima is reversed between "\emph{V}" and "\emph{W}", i.e. the average
distances of quasi-saddles from equilibrium configurations are practically
temperature independent while those of minima strongly decrease with
decreasing temperature. This kind of behaviour can be expected because the
potential energy of the starting equilibrium configurations is known to be
nearly proportional to temperature\cite{PRL85-5356} and configurations that
are not very "distant" from equilibrium are reasonably dependent on temperature.%

\begin{figure}
[ptb]
\begin{center}
\includegraphics[
]%
{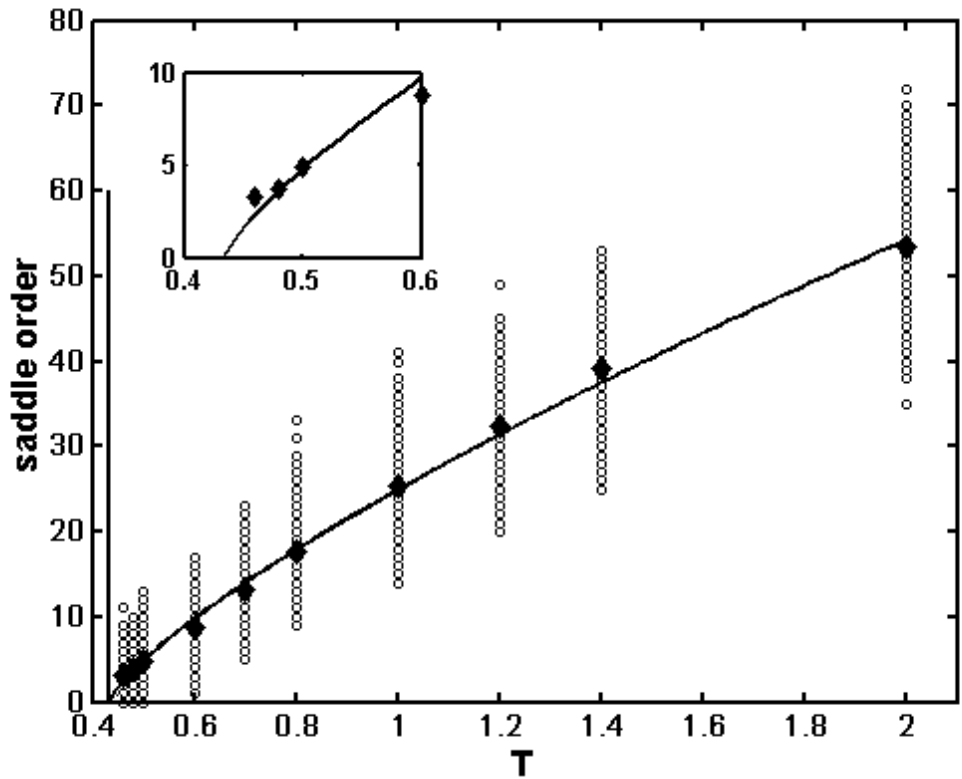}%
\caption{Saddle order (number of negative eigenvalues) for each quasi-saddle
configuration at different temperatures. Full diamonds represent average
values. Solid line is a power law fit to average values: const (T-T$_{c}%
$)$^{\gamma}$ with const $\simeq38.5$, T$_{c}$ $\simeq0.435$, $\gamma
\simeq0.77$. A solid vertical line is drawn at T$_{c}$. In the inset a blow-up
of the lowest temperature data is shown.}%
\label{ANeigW_T}%
\end{center}
\end{figure}

In figure \ref{ANeigW_T} we plot the number of negative eigenvalues of the
Hessian matrix at quasi-saddle points versus temperature. This number goes
fast to zero at the critical temperature, i.e. the temperature of structural
arrest, as already found by a similar analysis performed in ref.
\cite{PRL85-5356, PRL85-5360}. A solid vertical line in the figure represent
the value of the mode coupling critical temperature\cite{MCT92} (T$_{c}%
\simeq0.435$) derived from the power law behaviour of the diffusion
constant\cite{Kob95}. The same value is obtained by a power law fitting of
saddle order versus temperature (the solid line in figure \ref{ANeigW_T}) .
This confirms the number of negative eigenvalues of quasi-saddles (or saddles)
to be strictly connected with the diffusion constant. In contrast, a negative
eigenvalue of an equilibrium configuration, i.e. a negative curvature of the
instantaneous potential energy, is not always connected with a path joining
two different minima, i.e. a diffusion path.

\subsection{Quasi-saddles characteristics}

\begin{figure}
[ptb]
\begin{center}
\includegraphics[
width=6.6591in
]%
{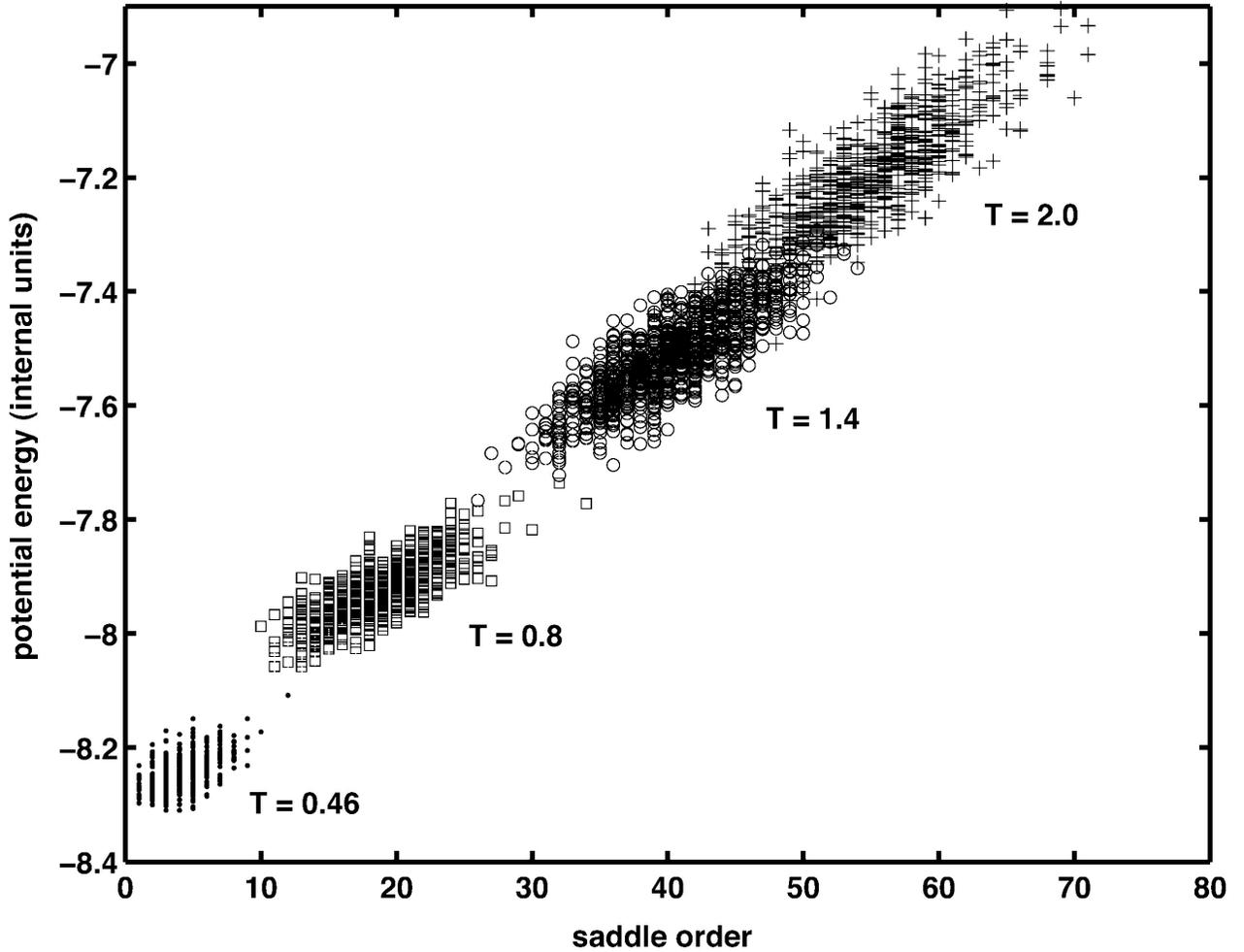}%
\caption{Potential energy of quasi-saddles versus saddle order (number of
negative eigenvalues) at some chosen temperatures: $T=0.46$ (dots),
$T=0.8$ (open squares), $T=1.4$ (open circles) and $T=2.0$ (plus signs). }%
\label{AphiW_eigNW}%
\end{center}
\end{figure}

As already noted in ref. \cite{PRL85-5360}, the quasi-saddles at different
temperatures explore different regions of the PEL. Therefore at low
temperatures we have low values of the potential energy of \ "\emph{W}"
together with low numbers of negative eigenvalues, as shown in figure
\ref{AphiW_eigNW}.%

\begin{figure}
[ptb]
\begin{center}
\includegraphics[
width=5.6196in
]%
{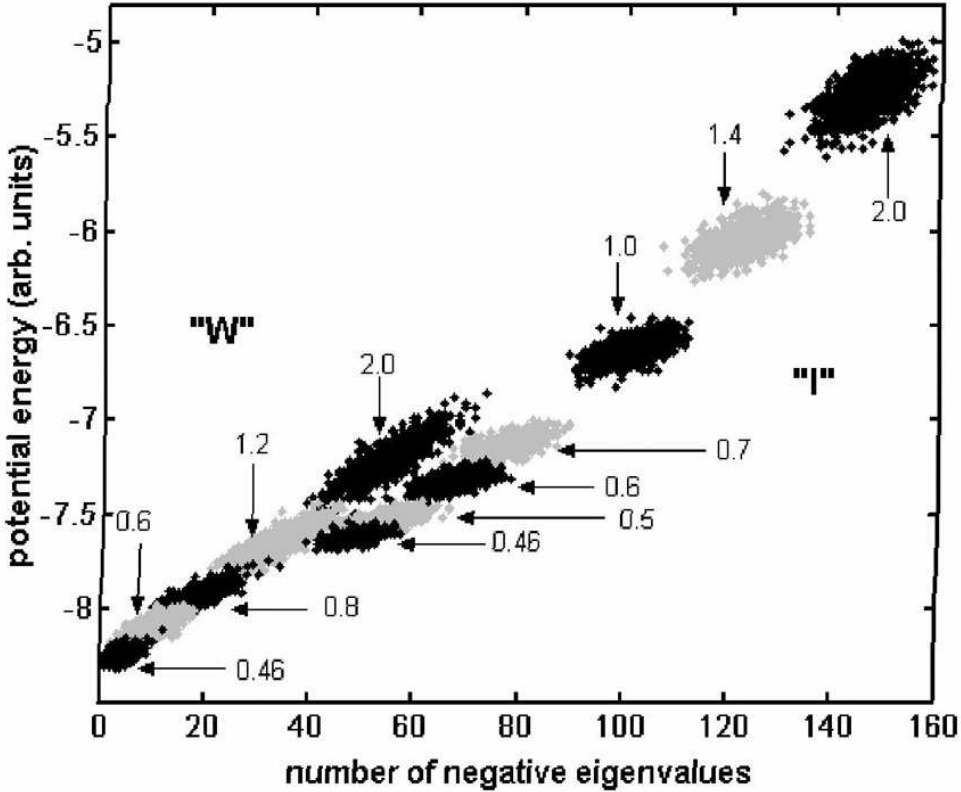}%
\caption{Potential energy of initial equilibrium configurations ("\emph{I}")
and quasi-saddles ("\emph{W}") versus number of negative eigenvalues at the
indicated temperatures. }%
\label{AphiIW_NeigIW}%
\end{center}
\end{figure}

We underline that the potential energy itself is not sufficient to determine
the number of negative eigenvalues of the Hessian matrix. Indeed even at \ the
same value of the potential energy the number of negative eigenvalues is
substantially different for quasi-saddles ("\emph{W}") and initial equilibrium
configurations ("\emph{I}"), as shown in figure \ref{AphiIW_NeigIW} where we
plotted the potential energy\ of "\emph{I}" and "\emph{W}" versus the number
of negative eigenvalues at different temperatures. From the values and
distribution in energy of "\emph{I}" and "\emph{W}" at the same temperature,
we can easily derive that the enormous decrease of diffusion (and increase in
viscosity) lowering the temperature in supercooled regime, is due essentially
to entropic barriers, i.e. to difficulties in reaching the saddle points, and
not to activated processes \cite{Sampoli03}.%

\subsection{Spectral densities of potential minima, saddles and initial
configurations}

To better understand the modifications of the PEL around "\emph{I}%
","\emph{W}","\emph{V}" configurations that occur lowering the temperature we
have study the frequency distribution of the vibrational states derived from
the Hessian matrix of the potential energy. To visualize all the spectral
features in a single plot, the eigenfrequencies of unstable directions (i.e.
the square roots of negative eigenvalues) are reported as negative frequencies.%

\begin{figure}
[ptb]
\begin{center}
\includegraphics[
width=15cm
]%
{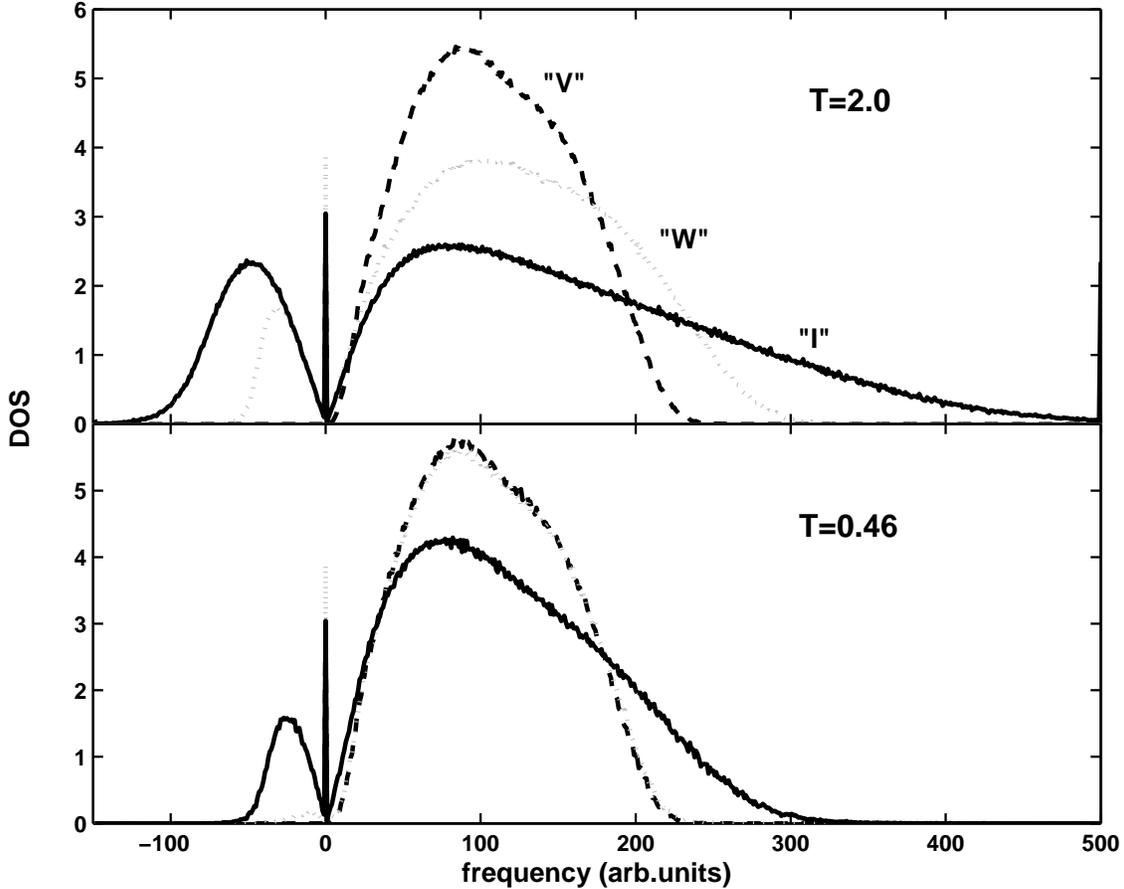}%
\caption{On the upper panel: the average density of states at high temperature
in the liquid regime from the normal mode analysis perfomed on the three
different configurations, \textquotedblright I\textquotedblright\ (solid
line),\textquotedblright W\textquotedblright\ (dotted grey line) and
\textquotedblright V\textquotedblright\ (dashed line). On the lower panel, the
same average DOS is shown near T$_{c}$ in the supercooled regime. Imaginary
frequencies (square root of negative eigenvalues) are reported as negative
frequencies. In the channel around zero of the histogram we found the three
zero frequency eigenvalues (connected to translational invariants); for
quasi-saddles another zero frequency eigenvalue is found due to the change of
curvature in one direction of the potential energy hypersurface (3 N = 768
dimensions). In the supercooled regime, approaching T$_{c}$, the
\textquotedblright W\textquotedblright\ DOS becomes more and more similar to
the \textquotedblright V\textquotedblright\ DOS.}%
\label{Ados1}%
\end{center}
\end{figure}

In figure \ref{Ados1} we reported the histogram\ of the average vibrational
density of states (DOS) at two extreme temperatures. From the figure we can
infer that by lowering the temperature all the three DOS will collapse into a
single one (when the glass will be confined to a single minimum). In the
supercooled regime, approaching T$_{c}$, the "\emph{W}" DOS from quasi-saddles
becomes more and more similar to the "\emph{V}" DOS from potential minima. At
T$=0.46$, the "\emph{W}" DOS is nearly coincident with the "\emph{V}" DOS
while the "\emph{I}" DOS of initial equilibrium configuration still remain
separated. We want to underline that in the channel of the histogram around
zero frequency, we found always the three eigenvalues connected to the three
translational invariants; for quasi-saddles (true-saddles apart) another zero
frequency eigenvalue is found due to the change of curvature in one direction
of the potential energy hypersurface. This is the reason why the zero channel
of the "\emph{W}" DOS has nearly four eigenvalues.%

\begin{figure}
[ptb]
\begin{center}
\includegraphics[
width=12cm
]%
{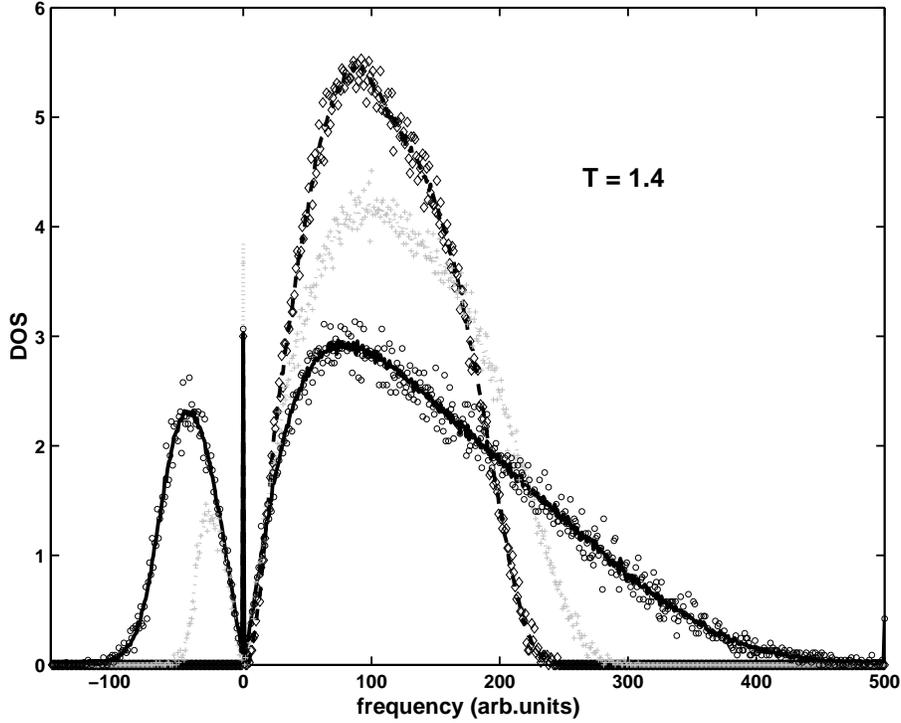}%
\caption{The average DOS at intermediate temperature (liquid regime) from the
normal mode analysis perfomed on the subset of different configurations giving
rise to true-saddles: "I" open circles, the true-saddle (grey crosses) and the
potential minimum (open diamonds) configurations. The average DOS from all the
investigated configurations at this temperature are reported as in Figure 2,
i.e. "I" (solid line),"W" (dotted grey line) and "V" (dashed line). Within the
statistical uncertainities the DOS are equal to those obtained from the
true-saddle subset. }%
\label{Ados2}%
\end{center}
\end{figure}

In figure \ref{Ados2}, the three DOS calculated from only the true-saddle
configurations, i.e. those configurations for which the forces at "\emph{W}"
points are really vanishing, are shown and compared with the DOS averaged over
all configurations. We want to stress that within the statistical
uncertainties the three DOS are not dependent on the chosen subset. This is
not obvious for "W" DOS and it is an important result of our study: at least
with respect to the distribution of vibrational frequency, true-saddles and
quasi-saddles share the same information.

\begin{figure}
[ptb]
\begin{center}
\includegraphics[
width=5.4276in
]%
{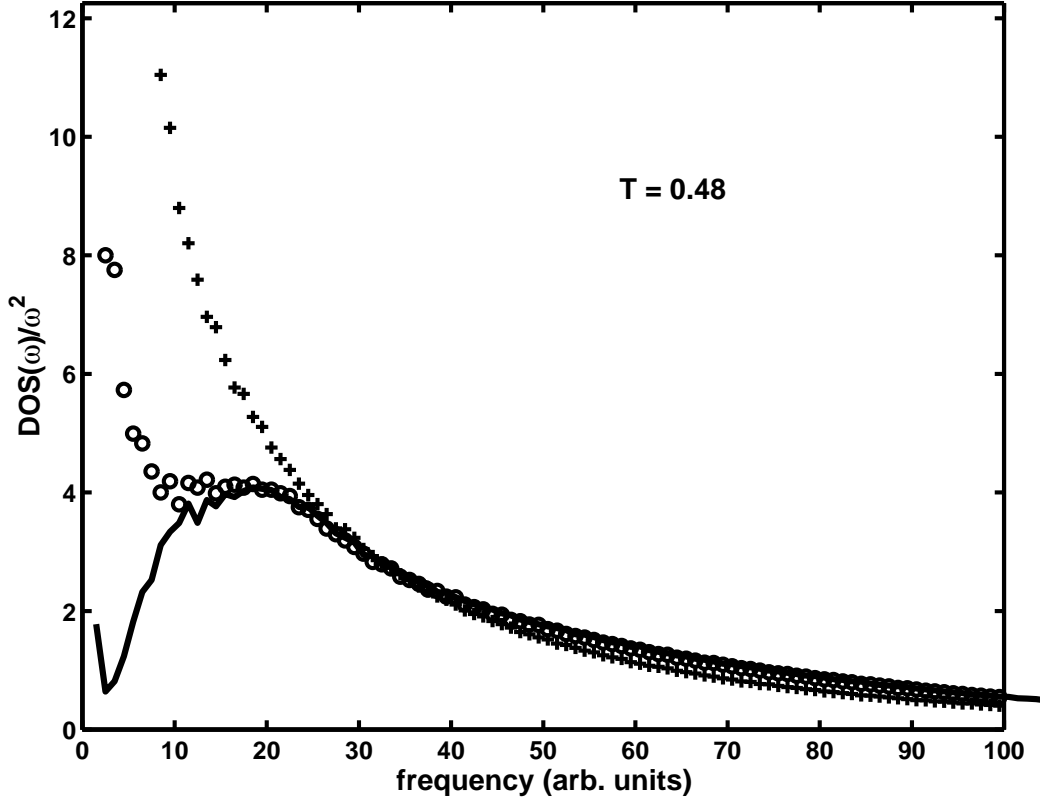}%
\caption{The average reduced DOS in the supercooled regime from the "I"
(crosses), "W" (circles) and "V" (solid line) configurations.}%
\label{Aboson_48}%
\end{center}
\end{figure}

Even if the "\emph{W}" and "\emph{V}" DOS seem to be very similar near T$_{c}$
(see lower panel of figure \ref{Ados1}) there are yet significant differences
at low frequency. In figure \ref{Aboson_48}, we reported the reduced density
of states (RDOS) at low frequencies for the three configuration at T$=0.48,$
i.e. the DOS divide by the frequency square. In the Debey solid approximation
the reduced density is a constant, while the well-known "Boson peak" is found
in real glasses. In the supercooled regime of figure \ref{Aboson_48}, a peak
is clearly visible in the "\emph{V}" RDOS, starts to appear in "\emph{W}"
RDOS, while is (yet) completely absent in the "\emph{I}" RDOS.%

\section{Conclusions}

We have investigated the dynamics of a model glass-forming liquids in terms of
its potential energy landscape by investigating numerous independent
configurations at different temperatures in the liquid and supercooled regime.
Potential energy minima, quasi-saddles and true-saddles are derived from
equilibrium configurations. The temperature behavior of potential energy,
saddle order, Cartesian distance, and vibrational density of states for
equilibrium, quasi-saddle and minimum configurations has been studied. In the
average the quasi-saddles lie on a more sliding hypersurface at low than high
temperatures. The number of true saddles is a decreasing function of
temperature, so it is important to establish that it is possible to extract
also information about the dynamics of the system from the quasi-saddles
configurations. It has been demonstrated that the vibrational density of
states is practically the same for true- and quasi-saddles. At different
temperatures, different regions of the PEL are statistically visited by the
system, due to the fact the accessible configuration volume is exponentially
growing with the temperature. The potential energy and the number of unstable
directions (i.e. negative eigenvalues) are statistically correlated both for
saddles and equilibrium configurations, but at the same potential energy
saddles and equilibrium configuration have different number of negative
eigenvalues, at least in the investigated liquid-supercooled region. An excess
density of states at low frequency is always present in the investigated
system, but a "Boson peak" like structure is clearly visible only in the DOS
derived from potential energy minima.

This work was realized with the financial support of MURST.


\begin{thebibliography}{99}                                                                                               %


\bibitem {Goldstein69}Goldstein M, 1969 \textit{J. Chem. Phys.} \textbf{51}, 3728

\bibitem {Keyes97}Keyes T, 1997 \textit{J. Chem. Phys.} \textbf{101}, 2921

\bibitem {Donati00}Donati C, Sciortino F, and Tartaglia P, 2000 \textit{Phys.
Rev. Lett.} \textbf{85}, 1464

\bibitem {Kurchan96}Kurchan J and Laloux L, 1996 \textit{J. Phys. A}
\textbf{29,} 1929

\bibitem {Cavagna98}Cavagna A, Giardina I, and Parisi G, 1998 \textit{Phys.
Rev. B }\textbf{57,} 11251

\bibitem {PRL85-5356}Angelani L, Di Leonardo R, Ruocco G, Scala A, and
Sciortino F, 2000 \textit{Phys. Rev. Lett.} \textbf{85}, 5356

\bibitem {PRL85-5360}Broderix K, Bhattacharya K K, Cavagna A, Zippelius A, and
Giardina I, 2000 \textit{Phys. Rev. Lett.} \textbf{85}, 5360

\bibitem {Shah01}Shah P, and Chakravarty C, 2001 \textit{J. Chem. Phys.}
\textbf{115, }8784

\bibitem {Cavagna01}Cavagna A, 2001 \textit{EuroPhys. Lett.} \textbf{53}, 490

\bibitem {Grigera02}Grigera T S, Cavagna A, Giardina I, and Parisi G, 2002
\textit{Phys. Rev. Lett.} \textbf{88}, 55502

\bibitem {Kob95}Kob W and Andersen H C, 1995 \textit{Phys. Rev. E
}\textbf{51,} 4626

\bibitem {phishift}The shift value is dependent slightly on temperature and
configuration ("$\emph{I}$","\emph{W}" or "\emph{V}", see text); the variation
is less than $\pm0.01$energy units.

\bibitem {Schroeder00}Schroeder T B, Sastry S, Dyre J C, and Glotzer S C, 2000
\textit{J. Chem. Phys.} \textbf{112, }9834

\bibitem {cm0108310}Doye J\ P K and Wales D J, 2002 \textit{J. Chem. Phys.}
\textbf{116, }3777

\bibitem {MS03}Sampoli M, to be published; a brief sketch is in the following.
All the tested programs for finding minima in a multidimensional space (768
dimension in our case) stick in some points when a tortuous deep valley is
encountered. Usually they decrease the step more and more, the search becomes
very slowly and possibly stops. The procedure can be tested easily in the case
of "\emph{V}" \ minima, because a potential minimum requires the sum of the
squared forces to be zero and the Hessian matrix to have all positive
eigenvalues (except three zeros). Different algorithms usually stick in
different points and the same algorithm with a larger step can be effective in
overcoming some critical situations. Therefore a complex flow chart with
different algorithms (steepest descent, Gauss-Newton, Levenberg-Marquardt,
preconditioned conjugate gradient, etc.) was used. In both "\emph{W}" and
"\emph{V}" minima we started with a steepest descent search.

\bibitem {Sastry99}Sastry S, Debenedetti P G, Stillinger F H, Schroeder T B,
Dyre J C, and Glotzer S C, 1999 \textit{Physica A} \textbf{270}, 301

\bibitem {MCT92}Goetze W and Sjoegren L, 1992 \textit{ Rep. Prog. Phys.}
\textbf{55}, 241

\bibitem {Sampoli03}Sampoli M, to be published. A brief summary is reported
here where, only the case of the lowest temperature (T$=0.46$) is examined.
The simulation was performed in NVE, so the fluctuations in "\emph{I}"
potential energy were equivalent to the fluctuations in kinetic energy. From
the "\emph{I}" potential energy values, the mean potential energy and its
standard deviation was estimated . From the mean temperature or the value of
total energy, the mean kinetic energy and its standard deviation (equal to the
standard devition of potential energy) was derived. The classical vibrational
potential energy was assumed to be equal to kinetic energy and the
contributions of diffusive modes were neglected at this temperature. The
combined fluctuations of potential and kinetic energy was estimated by
considering the two sources as statistically independent. If no activated
process is necessary to reach a saddle point, the saddle point energies must
be in the usual fluctuation range . With a fluctuation of two standard
deviations, $71.6\%$ of saddle points can be reached and $95.2\%$ with three
standard deviations. It must be noted that the microcanonical ensemble was
used to calculate this estimate, whereas also the fluctuations in total energy
must be taken into account in real systems. That further favors an essential
entropy barrier explanation. Large contributions from activated processes are
probably confined to low temperature glassy phases.
\end{thebibliography}
\end{document}